\documentclass[letterpaper, 10 pt, conference]{ieeeconf}  % Comment this line out if you need a4paper

\IEEEoverridecommandlockouts                              % This command is only needed if 
\usepackage{amsmath} % assumes amsmath package installed 
\usepackage{amssymb}  % assumes amsmath package installed
\usepackage{siunitx}
\usepackage[linesnumbered,ruled,vlined]{algorithm2e}
\usepackage{mathtools}
\usepackage{url}

\usepackage{pgfplots, pgfplotstable}
\usetikzlibrary{pgfplots.groupplots}
\pgfplotsset{compat = 1.3}

\newcommand{\R}{\mathbb{R}}

\makeatletter
\newcommand{\thickhline}{%
    \noalign {\ifnum 0=`}\fi \hrule height 2pt
    \futurelet \reserved@a \@xhline
}
\newcolumntype{"}{@{\hskip\tabcolsep\vrule width 1pt\hskip\tabcolsep}}
\makeatother

\title{\LARGE \bf
A real-time GP based MPC for quadcopters with unknown disturbances%Preparation of Papers for IEEE Sponsored Conferences \& Symposia*
}

\author{Niklas Schmid$^{1}$, Jonas Gruner$^{2}$, Hossam S. Abbas$^{2}$ and Philipp Rostalski$^{2}$% <-this % stops a space
\thanks{*Research funding: This work was funded by the German Federal Ministry of Education and Research under grant no. 16KIS1027 and the German Research Foundation (DFG), project number 419290163}. % Conflict of interest: Authors state no conflict of interest.}% <-this % stops a space
\thanks{$^{1}$N. Schmid, is with ETH Zürich – Automatic Control Laboratory, %\textcolor{orange}{ Physikstrasse 3, 8092 }
Zürich, Switzerland. Email:
        {\tt\small nikschmid@ethz.ch}}%
\thanks{$^{2}$J. Gruner, H. S. Abbas and P. Rostalski are with Universität zu Lübeck – Institute for Electrical Engineering in Medicine, 
Lübeck, Germany. Emails:
        {\tt\small \{j.gruner}, {\tt\small h.abbas}, 
        {\tt\small philipp.rostalski\}@uni-luebeck.de}}%
}

\begin{document}

\maketitle
\thispagestyle{empty}
\pagestyle{empty}

%TODOS:
% - mention jmavsim parameters?

%%%%%%%%%%%%%%%%%%%%%%%%%%%%%%%%%%%%%%%%%%%%%%%%%%%%%%%%%%%%%%%%%%%%%%%%%%%%%%%%
\begin{abstract}
%Model Predictive Control (MPC) \textcolor{orange}{is a modern control approach, which}
%one of the most important modern research fields in control theory due to its versatility and performance. However, since MPC} 
%relies on precise predictions of the system's state based on a model of the system's dynamics. \textcolor{green}{Thus,} the control performance significantly decreases if unknown disturbances affect the system. As a solution, 
Gaussian Process (GP) regressions have proven to be a valuable tool to %model and 
predict %prediction errors caused by 
disturbances and model mismatches and incorporate this information into a Model Predictive Control (MPC) prediction. Unfortunately, the computational complexity of  inference and learning on classical GPs scales cubically, 
%\textcolor{orange}{with $\mathcal{O}(n^3)$, where $n$ is the number of data points}, 
which is intractable for real-time applications. %classical GPs either rely on preselected, informative data and are trained offline, or they rely on large data sets of $n$ data points. Since the computational complexity of inference and learning on GPs scales with $\mathcal{O}(n^3)$ th later is intractable for real-time applications. 
Thus GPs are commonly trained offline, which is not suited for learning  disturbances as their  dynamics may vary with time.
%While it is often sufficient to collect data about modelling errors and train the corresponding GPs offline, this is not the case for disturbances whose dynamics may vary with time. 
%A new approach reformulates GPs as state-space systems and allows to pursue inference and learning with a computational complexity of $\mathcal{O}(n)$, which makes them attractive for real-time applications. 
 Recently, state-space formulation of GPs has been introduced, 
 %\textcolor{orange}{which allows to pursue the} 
 allowing inference and learning with %\textcolor{orange}{a} 
 linear computational complexity. %\textcolor{orange}{of $\mathcal{O}(n)$}. %This work presents a framework that allows to apply GPs in this state space formulation online to learn disturbance dynamics on a quadcopter system in the order of milliseconds. 
 %\textcolor{red}{In this paper, we present a framework that allows
 %online learning based on the state-space formulation of GPs to learn  disturbance dynamics on  %quadcopters, which can be executed  within milliseconds.} 
 %\textcolor{orange}{In this paper, we present} 
 This paper presents a framework that %\textcolor{orange}{allows}
 enables
 online learning of disturbance dynamics on quadcopters, which can be executed within milliseconds using a state-space formulation of GPs. 
 The obtained disturbance predictions are combined with MPC leading to a significant performance increase in simulations with jMAVSim. The %computation times are 
 computational burden is evaluated on a Raspberry Pi 4 B to prove the real-time applicability. % of the proposed approach. 
 
%This electronic document is a ÒliveÓ template. The various components of your paper [title, text, heads, etc.] are already defined on the style sheet, as illustrated by the portions given in this document. 

\end{abstract}

%%%%%%%%%%%%%%%%%%%%%%%%%%%%%%%%%%%%%%%%%%%%%%%%%%%%%%%%%%%%%%%%%%%%%%%%%%%%%%%%
\section{INTRODUCTION}

In recent years, unmanned aerial vehicles (UAVs) have increased in popularity due to their various possible applications, low cost, and variability. A common challenge in many UAV applications includes limitations in flight time due to a constrained battery capacity. A tethering system is a possible means to overcome this issue. However, such a tether creates an additional load and makes the UAV more susceptible to disturbances due to wind. 

The commonly used quadcopter setup for a UAV has 6 degrees of freedom (DoF). With four rotors, it is, therefore, an underactuated system, making it a good test platform for modern control approaches. A quadcopter is often controlled by a cascaded PID;
the use of MPC for quadcopters is also well established (cf. \cite{Islam2017}). However, most of these approaches cannot cope efficiently with exogenous disturbances due to the wind or a tether. There are different approaches to include disturbances in the control loop, such as disturbance rejection and backstepping (cf. \cite{Bannwarth2016}). This work proposes a more general approach by modeling disturbances as latent forces and moments acting on the quadcopter, which can be predicted using advanced machine learning techniques.

%\textcolor{red}{Learning-based techniques have attracted the control systems community for a long time \cite{Ko16}.}
In recent years, \emph{Gaussian process} (GP) regression models, which are nonparametric kernel-based probabilistic models \cite{RaWi05},  have demonstrated significant potential for learning control-oriented models, \cite{Sarkka2019}, with the ability to incorporate prior knowledge about the data through the choice of a \emph{kernel}. As a Bayesian framework, the strength of a GP regression is its ability to estimate the prediction uncertainty, which can be used for robustifying the control systems. 
%A GP distributed function can be given as \cite{RaWi05} $g(v)\sim{\mathcal{GP}}(\mu(v;\theta),\kappa(v,v^\prime;\theta))$, where $\mu(\cdot)$ is a mean function, $\kappa(\cdot,\cdot)$ is a covariance function (kernel) and $\theta$ is a so-called hyperparameter vector, which is usually determined (learned) 
%using maximum likelihood optimization \cite{RaWi05} given a set of data.
%When the function $g(\cdot)$ is evaluated on a finite set of inputs $v_1,v_2,\cdots$ the \emph{prior distribution} of the corresponding function values is given by a joint multivariate Gaussian distribution.
%Specifying a GP prior on $g(\cdot)$ via  $\mu(\cdot)$ and $\kappa(\cdot,\cdot)$ using  prior knowledge about $g(\cdot)$, then, conditioning on the data results in a predictive \emph{posterior distribution} at a new data point, which is Gaussian as well. 

%\textcolor{red}{The next two paragraphs are basically the same?}

In the context of MPC, it has been demonstrated in, e.g., \cite{HeKaZe20, GrMaAbRo19, MaGrRoAb21}, that GPs can approximate disturbances, nonlinearities, and model mismatches, where their information provided by the GPs can be exploited to enhance the MPC  performance. Moreover, the GPs probabilistic
descriptions have been leveraged  to propagate uncertainties of the estimates over the MPC prediction horizon, yielding a low conservative mechanism for robustification in terms of \emph{chance constraints}, which  casts the problem in a \emph{stochastic MPC}  framework \cite{HeKaZe20}.

This paper's contributions can be summarized as follows: We prove the feasibility and instruct the implementation of a real-time capable GP-based MPC for a quadcopter under disturbances using simulation. It is therefore assumed that the disturbances underlie some unknown dynamics. The idea is that if one uses GPs to mimic the underlying disturbance dynamics, then present and anticipated future disturbances can be inferred based on recorded past data. The predictions can then be used by MPC to counteract  these unknown disturbances. Furthermore, a cautious control is achieved by formulating a chance-constrained MPC problem using the uncertainties provided by the GP model, as proposed in \cite{HeKaZe20}.

\section{BACKGROUND}
\label{sec:background}
This section covers the required basics needed for this work. These include  Model Predictive Control (cf. \cite{Islam2017}), Gaussian processes (cf. \cite{RaWi05}) as well as the LTI representation of GPs with stationary kernels (cf. \cite{Hartikainen2010}).
\subsection{Model Predictive Control}
Assume a discrete-time nonlinear system given by
\begin{equation}\label{e:truesystem}
    x_{k+1} = f(x_k, u_k) 
\end{equation}
where $x_k\in\mathbb{R}^{n_x}$ is the state and $u_k\in\mathbb{R}^{n_u}$ the input of the system at time step $k$.  
Model Predictive Control builds upon a model $\hat{f}(x_k, u_k)$ of the system dynamics \eqref{e:truesystem}. %in $f(x_k, u_k)$.
Based on this model, the trajectory of the systems' state can be predicted in terms of the input and initial state. Thus,  the goal is to find 
\begin{align}
u^\ast(x_{t|t})=  && &  \underset{u_0,\cdots,u_{N-1}}{\mathrm{argmin}} l_f(x_{t+N|t}) + \sum_{k=0}^{N-1} l(x_{t+k|t},u_{t+k|t}) \nonumber\\
\text{s.t.}&& & x_{t+k+1|t}=\hat{f}(x_{t+k|t},u_{t+k|t})\nonumber\\
&&       & x_{t+k|t} \in\mathcal{X},\forall k=1,\cdots, N
   				   \label{eq_nominalMpcProblem}\\
&&       & u_{t+k|t} \in\mathcal{U},\forall k=0,1,\cdots, N-1 \nonumber
\end{align}
where $x_{t+k+1|t}$ is the prediction of $x_{t+k+1}$ at time $t$, $l(x_{t+k|t},u_{t+k|t})$ a stage cost function penalizing the state and input over the prediction horizon $N$, $l_f(x_{t+N|t})$  approximates the cost of regulating the state $x_{t+N|t}$ from time step $N$ to infinity and 
\begin{equation}\label{eq:constraints}
    \begin{aligned}
     \hspace{-2mm}   \mathcal{X}\!\coloneqq\!\{x\in\R^{n_x}|H^xx\leq h^x\},\
        \mathcal{U}\!\coloneqq\!\{u\in\R^{n_u}|H^uu\leq h^u\}
    \end{aligned}
\end{equation}
 are lower and upper bounds for the state and input, respectively, where $H^x\in\R^{n_{xc}\times n_x}$, $h^x\in\R^{n_{xc}}$, $H_u\in\R^{n_{uc}\times n_u}$, $h^u\in\nobreak\R^{n_{uc}}$ for $n_{xc}$ state and $n_{uc}$ input constraints. 
%\begin{subequations}
%\begin{alignat*}{4}
%& u^\ast(x_{t|t})= 	    \quad & \underset{x_k,u_k}{\mathrm{argmin}}&& & \sum_{k=0}^\infty l(x_{t+k|t},u_{t+k|t}) \\
%&     &\text{s.t.}&&     \quad  & x_{t+k+1|t}=\hat{f}(x_{t+k|t},u_{t+k|t})\\
%&     & 				   &&       & x_{t+k|t} \in\mathbb{X},\forall k\in \mathcal{N}\\
%&      & 				   &&       & u_{t+k|t} \in\mathbb{U},\forall k\in \mathcal{N}
%\end{alignat*}
%\end{subequations}
%here $x_{t+k+1|t}$ is the prediction of $x_{t+k+1}$ at time $t$, $l(x_{t+k|t},u_{t+k|t})$ a cost function penalizing the state and input of the system and $\mathcal{X}, \mathcal{U}$ polytopic constraints for the states and inputs.

\subsection{Gaussian Processes}
\label{section_GPsIntro}
Assume an unknown continuous-time function $g(t_i)$ with inputs $t_i$ and noisy outputs $y_i$ where $v_i$ denotes white Gaussian measurement noise such that
\begin{equation*}
  y_i = g(t_i) + v_i , \quad v_i\sim\mathcal{N}(0, \sigma_n^2),
\end{equation*} 
 where $t$ in this work will denote time.
 %can mean a number of things. However, throughout this paper it will denote time. 
 Gaussian Process Regression is concerned with the task of learning the function $g(t_i)$ such that predictions of the output $y_{n+1}$ can be made based on an input $t_{n+1}$ and a training set $(t_1, y_1),\dots,(t_n,y_n)$.

\begin{sloppypar}
Therefore it is assumed that the output data $y_{1:n+1}$ forms a multivariate normal distribution such that $y_{1:n+1}\sim \nobreak\mathcal{N}(\mu,K)$, where $\mu\in\R^{n+1}$ and the entries of $K\in\R^{n+1,n+1}$ at any element $K_{i,j}$ are given by the kernel function $k(t_i,t_j)$; the sequence $y_{1:n+1}$ here denotes a column vector of length $n+1$. Thus the correlation of any two $y_i$ and $y_j$ is given by the kernel function and the input data $t_i, t_j$. A prediction of the unknown output $y_{n+1}$ is obtained from the multivariate normal distribution by marginalizing over $y_{1:n}$. Since %one thus obtains
normally distributed predictions  $y_{n+1}$ are obtained for arbitrary input values $t_{n+1}$ this procedure yields a distributed function $y_{n+1}=\hat{g}(t_{n+1})$ called a Gaussian Process. The mean $\mu$ defines a bias and is often assumed zero.
\end{sloppypar}
% is it intentional that the indices jump up and down? JG - Maybe they're happy. Fixed it. NS

The kernel which determines the data points' correlation is usually parameterized by a set of hyperparameters. These hyperparameters are fitted to the data such that the kernel function yields a most likely description of the data points true correlation. More formally, the optimal hyperparameters are determined by
\begin{equation}\label{e:opt_hyper}
    \max_{\theta} p(\theta|t_{1:n}, y_{1:n}),
\end{equation}
where $\theta$ is a vector of the sought hyperparameters.
In practice, one may use gradient descent methods to maximise the (log) likelihood of the optimization problem \eqref{e:opt_hyper} and update the GP hyperparameters iteratively as
\begin{equation}
    \theta_{n+1} = \theta_n + \eta \frac{dp(\theta|t_{1:n}, y_{1:n})}{d\theta}
    \label{eq_hyperparametersUpdate}
\end{equation}
where $\eta$ denotes a learning rate. % This is denoted as learning. 

The most widely used kernel, the squared exponential kernel
%\begin{equation}
$k(t,t') = \sigma_m^2\exp(-(t-t')^2/l^2)$
%\label{label_eq_basicsGPSEK}
%\end{equation}
with hyperparameters $l$ and $\sigma_m^2$, assumes a high correlation when the data points $t$ and $t^\prime$
are close, thus, the corresponding output data points are likely to be similar. Therefore, it is often used 
for learning smooth functions.
%assumes a high correlation for data points whose inputs are have a small absolute difference. Thus output data points are likely similar if the respective input data has a small absolute difference, which is identical to assuming smoothness of the underlying function $g(t)$. 
By varying the width of the squared exponential kernel, i.e., $l$, %which is one of its hyperparameters, 
one can adjust the input range over which the output data is assumed to be similar.   % and thus adjust the anticipated  smoothness of the function $g(t)$. %%\pgfplotsset{%
%    width=.3\textwidth,
%    height=.2\textwidth
%}
%\begin{figure}
%\centering
%\begin{tikzpicture}
% %1/e=0.367
%\begin{axis}[ytick={1
%},
%yticklabels={$\sigma_m^2$
%},
%xtick={\empty
%},
%xticklabels={\empty
%},
%axis x line=bottom,axis y line=left,xlabel={$t-t'$}, ylabel={$k(t-t')$}, axis %lines=middle,
%    xmin = -3, xmax = 3,
%    ymin = -0.1, ymax = 1.2]
%    \addplot[
%        domain = -5:5,
%        samples = 200,
%    ] {exp(-x^2)};
%    \draw [<->, thick] (-1, 0.367) -- (1,0.367);
%		\draw[] (0.5,0.367) node[above] {$l^2$};
%\end{axis}
%\end{tikzpicture}
%\caption{The parametrization of the squared exponential kernel.}
%\label{img_squaredExpKernel}
%\end{figure}

The inference, or more precisely, marginalization in GPs, can be done in closed form as well as the calculation of the gradient in \eqref{eq_hyperparametersUpdate}. %({\color{red} for completeness it is good to write these equations})  -> See commented
%\begin{equation}
%y^*|t_{1:n},y_{1:n},t^*  \sim \mathcal{N}(\bar{y}^*, cov(y^*))
%\end{equation} 
%with
%\begin{equation}
%\begin{aligned}
%\bar{y}^* &= K^*(K+\sigma^2_nI)^{-1}\textbf{y}, \\
%cov(y^*) &= K^* - K^*(K+\sigma^2_nI)^{-1}K^{*T}.
%\end{aligned}
%\label{label_eq_inferenceClassicalGPs}
%\end{equation}
%\begin{align}
%\frac{\delta}{\delta \theta_j}ln(p(\textbf{y}|\theta, \textbf{t})) = %\frac{1}{2}\textbf{y}^TK^{-1}\frac{\delta K}{\delta \theta_j}K^{-1}\textbf{y} - %\frac{1}{2}tr\left(K^{-1}\frac{\delta K}{\delta \theta_j}\right).
%\label{label_eq_learningOnClassicalGPsGradient}
%\end{align}
However, both require the inversion of the covariance matrix $K$, which yields a computational complexity of $\mathcal{O}(n^3)$. Thus, with increasing data, GPs become intractable for real-time applications. 

A solution to this problem was proposed % by Hartikainen et al.
in \cite{Hartikainen2010} by reformulating various GPs with stationary kernels $k(t_j,t_i)=k(\tau)$, where $\tau=t_j-t_i$, as LTI-systems on which inference and learning are carried out  using Kalman-Filtering and Rauch-Tung-Striebel-Smoothing with a computational complexity of $\mathcal{O}(n)$. 

The transformation is done as follows: According to the Wiener–Khinchin theorem the spectral density $S(i\omega)$ of $\hat{g}(t)$ can be obtained by the Fourier-Transform of the stationary kernel function $S(i\omega)=\mathcal{F}\{k(\tau)\}$. Assuming a finite spectrum of $S(i\omega)$, a consecutive spectral factorization $\mathcal{F}\{k(\tau)\}=H(i\omega)qH(i\omega)$ yields a system with the transfer function $H(i\omega)$ driven by white Gaussian noise with spectral density $q$ which yields an output with the same spectral density as $\hat{g}(t)$. Thus, the  system driven by the noise   can be written as a continuous LTI-system with a discrete output
\begin{equation}\label{eq:CTGPSS}
    \begin{aligned}
    \dot{z}(t) &= Fz(t) + Lw(t), \quad
    \hat{y}(t_i) = Hz(t_i) + \hat{v}(t_i) 
\end{aligned}
\end{equation}
where $z\in\R^{n_z}$ denotes the state vector of the GP state-space model, $w(t)\sim\mathcal{N}(0,q)$,  $\hat{v}(t_i)\sim\mathcal{N}(0,\hat{\sigma}^2_n)$ represents measurement noise and the matrices $F,L,H$ are of appropriate dimension. The system yields a prediction $\hat{y}(t_i)$ for arbitrary inputs $t_i$ based on the state $z(t_i)$. 

For the inference, the system is discretized at a sampling distance $\Delta T$. Given a record of past outputs $y_{1:n}$, the state $z_{n+1}\sim({m^z_{n+1},\Sigma^z_{n+1}})$ can be inferred by Kalman-Filtering, which yields a prediction of the output \mbox{$y_{n+1} \sim \mathcal{N}(Cm^z_{n+1},C\Sigma^z_{n+1}C^\top + \hat{\sigma}^2_n)$}. For the learning, the gradient in \eqref{eq_hyperparametersUpdate} can be obtained efficiently by refactoring results of the inference \cite{Hartikainen2010}. Finally, it is worth mentioning that  $\mathcal{F}\{k(\tau)\}$ does not always yield the proposed rational form for arbitrary kernels, e.g., the squared exponential kernel. In such cases, it can be approximated to a desired order  using a
Taylor series or a Padé approximation, see \cite{Sarkka2019} for more details.
%\textcolor{red}{Should we add citations in the introduction of GPs and MPC?} \textcolor{orange}{Done!} 
%. It can then by obtained by a Taylor approximation of desired order \cite{Sarkka2019}.

% TODO CLEAR DIFFERENTIATION GP AND UNDERLYING FUNCTION...
\section{Modeling}
\label{sec:modeling}
The quadcopter position is denoted as $p\in\R^3$ and velocity $\dot{p}\in\R^3$ in an inertial frame which is spanned by unit vectors $e_{\hat{x}}, e_{\hat{y}}, e_{\hat{z}}\in\R^3$. Furthermore, the attitude between the inertial frame and the body frame, which is aligned with the quadcopter axes along the unit vectors $e_{x}, e_{y}, e_{z}\in\R^3$, is given by the Euler angles $\Phi\in\R^3$ and attitude rate $\dot{\Phi}\in\R^3$ in the roll, pitch and yaw directions. The matrix $R_{rot}\in\R^{3\times 3}$ denotes the rotation matrix from the body frame to the inertial frame.

The quadcopter system is modeled as a rigid body with three rotational and three translational DoF and dynamics
$\ddot{p} = F/ m,$ $M= J\dot{\omega} + \omega \times J\omega$ \begin{mbox}where $m\in\R$ is the quadcopter mass, $g=\SI{9.81}{\meter\per\second\squared}$, $\omega\in\nobreak\R^3$ denotes the angular rate of the quadcopter in the body frame, $F\in\R^3$ the sum of all forces and $M\in\R^3$ the sum of all moments which act on the quadcopter body.
\end{mbox}

The force $F$ is a summed effect of gravity, the propeller actuation, and aerodynamics, which, for simplicity, are assumed to scale linearly with the quadcopter velocity such that
$F = R_{rot} \sum_i F_i e_{z} - mge_{\hat{z}} - k_D\dot{p},$ 
where $F_i\in\R$ denotes the thrust generated by a single propeller $i$ and $k_D\in\R$ the drag force coefficient.

The moment $M=\begin{pmatrix}
\tau_x & \tau_y & \tau_z
\end{pmatrix}\in\R^3$ also stems from the actuation of the quadcopter's propellers and allows the quadcopter to roll, pitch and yaw. Moments caused by blade flapping (cf. \cite{Waslander2009}) are neglected in this model for simplicity.

In order to derive a linear model for the quadcopter a state $x = \begin{pmatrix} p^\top & \dot{p}^\top & \Phi^\top & \dot{\Phi}^\top \end{pmatrix}^\top\in\R^{12}$ and normalized inputs $u=\begin{pmatrix} \frac{T}{T_{max}} & \frac{\tau_x}{\tau_{x,max}} & \frac{\tau_x}{\tau_{y,max}} & \frac{\tau_x}{\tau_{z,max}} \end{pmatrix}^\top\in\R^4$ are defined, where $T[\SI{}{\newton}]$ is the thrust, $\tau_x[\SI{}{\newton\meter}],\tau_y[\SI{}{\newton\meter}],\tau_z[\SI{}{\newton\meter}]$ the longitudinal, lateral  and vertical torques and $T_{max},\tau_{x,max},\tau_{y,max},\tau_{z,max}$ the respective maximum values, such that $\frac{T}{T_{max}}\in[0,1],\frac{\tau_x}{\tau_{x,max}}\in[-1,1],\frac{\tau_y}{\tau_{y,max}}\in[-1,1],\frac{\tau_z}{\tau_{z,max}}\in[-1,1]$. The state and input dimensionality are denoted with $n_x=12$ and $n_u=4$, respectively. 

Linearizing the nonlinear dynamics around the hovering condition  $\dot{x}_s =0$ and $u_s = \begin{pmatrix} \frac{mg}{T_{max}} & 0 & 0 & 0 \end{pmatrix}^{\top}$ yields the linear model
\begin{align}
    \dot{x} &\approx 
    \begin{pmatrix} 
    0 & I & 0 & 0\\
    0 & -\frac{k_D}{m}I & A_1 & 0 \\
    0 & 0 & 0 & I \\
    0 & 0 & 0 & 0
    \end{pmatrix}x
    + \begin{pmatrix} 
    0 \\
    B_1  \\
    0 \\
    B_2
    \end{pmatrix}(u-u_s),
    \label{eq_linear_model_quadcopter}
\intertext{where}
    A_1 &= \begin{pmatrix} 
    0 & g & 0 \\
    -g & 0 & 0 \\
    0 & 0 & 0 
    \end{pmatrix}, \ 
    B_1 = \begin{pmatrix} 
    0 & 0 & 0 & 0\\
    0 & 0 & 0 & 0\\
    \frac{1}{m}T_{max} & 0 & 0 & 0
    \end{pmatrix}, \nonumber \\
    B_2 &= \begin{pmatrix} 
    0 & \frac{1}{I_{xx}}\tau_{x,max} & 0 & 0\\
    0 & 0 & \frac{1}{I_{yy}}\tau_{y,max}  & 0\\
    0 & 0 & 0 & \frac{1}{I_{zz}}\tau_{z,max}  \nonumber 
    \end{pmatrix}.
\end{align}

%The moment of inertia $J$ is obtained from the bifilar pendulum experiment (tdocite). The aerodynamic drag coefficient $k_D$ is obtained by flying in one direction at a constant speed and height. The respective attitude angle yields the horizontal thrust, since the vertical thrust component equals $mg$ and the propellers' thrust vectors is aligned with the body axis $e_z$.
%The constraints of the MPC require the actuation limits. The maximum thrust and torque in the roll and pitch direction are obtained from the maximum thrust of a single propeller, which can be approximated by the necessary throttle during hovering. Therefore, flight tests have been made using a Piwhawk 4 quadcopter which has normalized input commands for the throttle and torques. ... The maximum vertical torque $\tau_{z,max}$ is obtained during flight tests by applying aggressive yaw commands to the quadcopter during hovering and logging the acceleration of the yaw-angle. Then $T_{\psi,max} = \arg\!\min_{\gamma}||\ddot{\psi}-\frac{\gamma}{I_{zz}}\tau_z||^2$. 
% ^
% |
% |
% |
% Is this part necessary? We do not use the quadcopter overall. It could be any kind of quadcopter. We could as well just state values I think? - NS

The wind and the tether will act as latent forces and moments on the quadcopter body. Assuming the tether is attached to the center of mass, we will only focus on latent forces. However, the proposed procedure can easily be extended to include latent moments. 

\section{Controller Design}
\label{sec:Controller_Design}
GPs are used here to predict the latent forces caused by the disturbances over the prediction horizon of the MPC. This information is then used to improve the MPC performance and guarantee constraint satisfaction up to probability levels $p_x$ and $p_u$ for the state and input, respectively. The control policy is thus to find an input sequence $u_{1:N}$ for the stochastic optimization problem
\begin{align}%{4}
 \min_{u_0,\cdots,u_{N-1}} && & E\left[\|x_{N|t}\|_P + \sum_{k=0}^{N-1} \|x_{t+k|t}\|_Q + \|u_{t+k|t}\|_R\right]   \nonumber \\
\text{s.t.}&& & x_{t+k+1|t}=\hat{f}(x_{t+k|t},u_{t+k|t},d_{t+k|t})  \nonumber\\
&& & {\rm Pr}(x_{t+k|t} \in\mathcal{X})\leq p_x,\forall k= 1,\dots,N 
   				   \label{eq_stochasticMpcProblem}\\ 
&& & {\rm Pr}(u_{t+k|t} \in\mathcal{U})\leq p_u,\forall k= 0,\dots,N-1  \nonumber
\end{align} 
where $E$ denotes the expected value, $Q=Q^\top\succeq 0$, $P=P^\top\succeq 0$ and $R=R^\top\succ 0$, the norm $\|x\|_A = x^\top Ax$ denotes the matrix norm and $d_{t+k|t}$ represents a prediction at time $t$ of the disturbance at time step $t+k$. While $Q$ and $R$ are used for tuning the MPC behavior, the matrix $P$ weights the terminal cost. Note that computing the above problem is intractable, therefore, simplifications will be introduced.

\subsection{Disturbance extraction}
\label{subsec_disturbanceExtraction}
The latent forces in all DoFs produce accelerations in the different directions  of the quadcopter body. In order to train GPs to learn the dynamics of these accelerations, it is first necessary to extract and record data of past accelerations due to the disturbances. This can be performed by comparing measured velocities $\dot{p}_{t|t}$ of the quadcopter with 
simulated velocities $\dot{\hat{p}}_{t|t}$ based on the proposed nonlinear quadcopter dynamics.
%the respective predictions $\dot{\hat{p}}_{t|t}$ which are obtained based on the proposed nonlinear quadcopter dynamics. 
Therefore, the accelerations in the different directions  can  be estimated by
\begin{equation*}
\ddot{p}_{d,i,t-1|t} = \frac{\dot{p}_{i,t|t}-\dot{\hat{p}}_{i,t|t}}{\Delta T}
\end{equation*}
where $i\in\{\hat{x},\hat{y},\hat{z}\}$ denotes the direction of the acceleration 
and $\Delta T$ the discretization length of the state prediction. % too unprecise?
The recorded accelerations $\ddot{p}_{d,i,0|t},\dots,\ddot{p}_{d,i,t-1|t}$ in all DoFs are assumed to be uncorrelated since they stem from unknown disturbance dynamics and since the translational dynamics of the quadcopter can be separated into the single DoFs. Thus, the inference and learning in each DoF can be handled independently by a single one-dimensional GP such that three GPs are assigned to predict the  accelerations associated with the disturbances. 
%caused by the disturbances This leads to three GPs trained to mimic the dynamics of the accelerations caused by the disturbances in a specific DoF and used to predict future accelerations associated with the disturbances in the same respective DoF. 
%Note that three GPs with one-dimensional inputs and outputs are computationally more attractive than a single GP with a three-dimensional input and output ({\color{red}   this sentence is confusing}). 

The three GPs to learn and predict  $\ddot{p}_{d,i}(t)=g_i(t)$ with $i\in\{\hat{x},\hat{y},\hat{z}\}$ will be used in their continuous-time state-space representation with the  deterministic formulation
\begin{equation}
\begin{aligned}
\dot{z}_i(t) &= F_{i}z_i(t), \quad
\hat{y}_i(t) &= H_{i}z_i(t).
\end{aligned}
\label{eq_GPLTIcont}
\end{equation}

\subsection{Predictor} %Deterministic GP based MPC}
To define the predictor of the MPC, we combine the model of the quadcopter system and the deterministic state-space models of the GPs \eqref{eq_GPLTIcont}, which results in  the augment model % the quadcopter system leading to 
\begin{align} 
\label{eq_augmentedStateSpaceEq} 
    \begin{pmatrix}
\dot{x}      \\
\dot{z}_{\hat{x}}\\
\dot{z}_{\hat{y}}\\
\dot{z}_{\hat{z}}
\end{pmatrix} &= \begin{pmatrix}
A & C_{\hat{x}}H_{\hat{x}} & C_{\hat{y}}H_{\hat{y}} & C_{\hat{z}}H_{\hat{z}} \\
0 & F_{\hat{x}} & 0 & 0 \\
0 & 0 &  F_{\hat{y}} & 0 \\
0 & 0 &  0 & F_{\hat{z}} 
\end{pmatrix}     \begin{pmatrix}
\dot{x}      \\
\dot{z}_{\hat{x}}\\
\dot{z}_{\hat{y}}\\
\dot{z}_{\hat{z}}
\end{pmatrix}
 + \begin{pmatrix}
B \\ 0 \\ 0 \\ 0
\end{pmatrix}
u 
\end{align}
where $A, B$ are obtained from the quadcopter model in \eqref{eq_linear_model_quadcopter} and
the matrices $C_{\hat{x}}, C_{\hat{y}}, C_{\hat{z}}\in\R^{n_x}$ are column vectors mapping the predicted accelerations by the three GPs into the respective dimension of the quadcopter's state. Then, discretizing the augmented model above yields the predictor model of the MPC, which  directly incorporates the GP's disturbance predictions during state propagation. In this work, we adopt the exact discretization approach \cite{GrMaAbRo19}.
Note that the discrete-time form of  \eqref{eq_augmentedStateSpaceEq} 
represents $\hat{f}(x_{t+k|t},u_{t+k|t},d_{t+k|t})$  in  \eqref{eq_stochasticMpcProblem}.

%The above state-space equation is discretized and used as the predictor model $\hat{f}(x_{t+k|t},u_{t+k|t},d_{t+k|t})$ of the MPC problem in eq. \ref{eq_stochasticMpcProblem}, which thus incorporates the GP's disturbance predictions during state propagation.

\subsection{Uncertainty propagation and constraints tightening} %Stochastic GP based MPC}
To incorporate the uncertainty of the GP predictions in the MPC formulation, we make use of some formulations  presented in \cite{HeKaZe20}. We consider the discrete-time representation of \eqref{eq:CTGPSS} %The predictions of the three GPs can be obtained by discretizing the continuous stochastic LTI system of the GPs 
as 
\begin{align*}
z_{i,t+k|t} &=\bar{F}_iz_{i,t+k-1|t} + \bar{L}_iw_{i,t+k-1|t} \\
\hat{y}_{i,t+k|t} &=\bar{H}_iz_{i,t+k|t} + \hat{v}_{i,t+k|t},
\end{align*}
where the variables $\bar{F}_i$, $\bar{H}_i$, $\bar{L}_i$, $w_{i,t+k|t}\sim\mathcal{N}(0,Q_i)$, $\hat{v}_{i,t+k|t}\sim\nobreak\mathcal{N}(0,\hat{\sigma}_{n,i}^2)$ are the discrete time representatives of the continuous time  variables in \eqref{eq:CTGPSS} and  $i\in\{\hat{x},\hat{y},\hat{z}\}$. Therefore, the mean value and the covariance of the state of the GPs can be propagated as follows:
\begin{equation}\label{eq:propagateGPstate}
    \begin{aligned}
     \bar{z}_{i,t+k|t} &=   \bar{F}_i\bar{z}_{i,t+k-1|t} \\
     \Sigma^{z}_{i,t+k|t}&=\bar{F}_i\Sigma^{z}_{i,t+k-1|t}\bar{F}^\top_i + \bar{L}_iQ_i\bar{L}^\top_i
    \end{aligned}
\end{equation}
and for the output of the GPs, they are
\begin{equation}\label{eq:propagateGPoutput}
    \begin{aligned}
  \bar{y}_{i,t+k|t} &=  \bar{H}_i\bar{z}_{i,t+k|t} \\
  \Sigma^d_{i,t+k|t}&=\bar{H}_i\Sigma^{z}_{i,t+k|t}\bar{H}^\top_i + \hat{\sigma}_{n,i}^2,
  \end{aligned}
\end{equation}
with  $i\in\{\hat{x},\hat{y},\hat{z}\}$.
%
%predict the  $z_{i,t+k|t}\sim\mathcal{N}(\bar{z}_{i,t+k|t},\Sigma^{z}_{i,t+k|t})$ %
%
%The expectation and uncertainty of the GP state thus propagates as 
%\begin{align}
%z_{i,t+k|t} &\sim\mathcal{N}\left(\bar{F}_i\bar{z}_{i,t+k-1|t}, \bar{F}_i\Sigma^{z}_{i,t+k-1|t}\bar{F}^\top_i + %\bar{L}_iQ_i\bar{L}^\top_i\right) \nonumber \\
%g_{i,t+k|t} &\sim\mathcal{N}(\bar{H}_i\bar{z}_{i,t+k|t}, \underbrace{\bar{H}_i\Sigma^{z}_{i,t+k|t}\bar{H}^\top_i %+ \Sigma^n_{i}}_{\Sigma^d_{i,t+k|t}}).
%\end{align} 

%\begin{mbox}
%The GP provides predictions as normally distributed variables.
The normal distribution of the GP predictions yields the quadcopter state stochastic  
%When propagating the quadcopter state over the prediction horizon the uncertainty of the GP predictions leads to an uncertainty of the quadcopter state 
such that $x_{t+k|t}\sim\nobreak\mathcal{N}(\bar{x}_{t+k|t}, \Sigma^{x}_{t+k|t})$, where
%\end{mbox}
 the quadcopter state evolves according to the nominal system in   \eqref{eq_linear_model_quadcopter} plus the additive disturbances $\Sigma^d_{i,t+k|t}$. Therefore, using \eqref{eq:propagateGPstate} and \eqref{eq:propagateGPoutput}, the covariance of the quadcopter state propagates as
%One would propagate the uncertainty of the quadcopter state via its nominal system in eq. \ref{eq_linear_model_quadcopter} plus the uncertainty of the additive disturbances $\Sigma^d_{i,t+k|t}$, such that
\begin{equation}\label{eq:state-uncertainty-propagation}
\Sigma^x_{t+k+1|t} = A\Sigma^x_{t+k|t}A^\top + \sum_{i\in\{\hat{x},\hat{y},\hat{z}\}}C_i\Sigma^d_{i,t+k|t}C_i^\top.
\end{equation} Since the quadcopter system is unstable and the MPC only stabilizes the nominal system based on its deterministic predictor model, the evolution of the covariance of the quadcopter state may diverge over the prediction horizon according to \eqref{eq:state-uncertainty-propagation}. %during this propagation.
To account for this effect, an LQR controller $K_{\infty}$ is used. % introduced to restrict the uncertainty.
The input then reads as
\begin{equation}
    u_{t+k|t} = \Bar{u}_{t+k|t} + K_{\infty}(\Bar{x}_{t+k|t} - x_{t+k|t}).
    \label{eq_dual_control_input}
\end{equation}
where $\Bar{u}_{t+k|t}$ is computed by the MPC. Then, the growth of the uncertainty will be restricted 
as %uncertainty of the state then propagates via the closed-loop system as
\begin{multline*}
\Sigma^x_{t+k+1|t} = (A-BK)\Sigma^x_{t+k|t}(A-BK)^\top \\  + \sum_{i\in\{\hat{x},\hat{y},\hat{z}\}} C_i\Sigma^d_{i,t+k|t}C_i^\top.
\end{multline*}
The feedback of the stochastic state in \eqref{eq_dual_control_input} %, the uncertainty of the state 
yields the control input stochastic with the distribution %, which can be computed as
\begin{equation*}
    u_{t+k|t} \sim \mathcal{N}(\bar{u}_{t+k|t}, K\Sigma^x_{t+k|t}K^\top).
\end{equation*}
The stochastic nature  of the state and input at any time step can be described by defining uncertainty regions $S^x_{t+k|t}$ and $S^u_{t+k|t}$, respectively, around their expected values within which their uncertain values lie up to some probability level.
%The uncertainty of the state and input at every time step is useful to define uncertainty regions $S^x_{t+k|t}$ around the expectation of the state in which the uncertain state will lie up to some probability level. 
These regions can then be subtracted from the original constraint sets $\mathcal{X}$ and $\mathcal{U}$ to obtain tightened sets $\tilde{\mathcal{X}}_{t+k|t} =\mathcal{X} \ominus S^x_{t+k|t}$ and $\tilde{\mathcal{U}}_{t+k|t}=\mathcal{U} \ominus S^u_{t+k|t}$, where $\ominus$ denotes the Pontryagin set difference, which are used to define guarantees on the satisfaction of the chance constraints in \eqref{eq_stochasticMpcProblem} \cite{HeKaZe20}. If the state/input lies within these tightened sets, then, up to the probability level, the quadcopter state/input will not exceed the original constraints despite the unknown disturbances. For the applied half-space constraints \eqref{eq:constraints}, the tightened constraint sets can be computed as  \cite{HeKaZe20}
\begin{align}
\label{eq_tightenedConstraints}
\begin{aligned}
\tilde{\mathcal{X}}_{t+k|t} %&= \mathcal{X} \ominus S^x_{t+k|t}  \\ 
&\!=\!\left\{ \!x_{t+k|t}| H^xx_{t+k|t}\! \leq\! l^x\right.  \\
&\qquad\qquad\left. - \!|H^x|\Phi^{-1}\!\left( \bar{p}\right)\sqrt{{\rm diag}(\Sigma^x_{t+k|t})} \right\} \\
\tilde{\mathcal{U}}_{t+k|t} %&= \mathcal{U} \ominus S^u_{t+k|t}  \\
&=\left\{ u_{t+k|t}| H^uu_{t+k|t} \leq l^u 
\right.  \\
&\qquad\qquad\left.-\! |H^u|\Phi^{-1}\left( \bar{p}\right)\sqrt{{\rm diag}(\Sigma^u_{t+k|t})} \right\}
\end{aligned}
\end{align}
where ${\rm diag}$ indicates a vector of the matrix diagonal elements, $\Phi^{-1}$ is the quantile function of a standard normal distribution, $\bar{p}=1-\left(\frac{1}{n_x}-\frac{p_x+1}{2n_x}\right)$ where $0\leq p_x\leq 1$ is some probability level and
$\Sigma^u_{t+k|t}=K\Sigma^x_{t+k|t}K^\top$,
%and $\mathcal{X}$, $\mathcal{U}$ represent the nominal MPC constraints for the state and input, respectively, 
see \cite{HeKaZe20} for more details. Finally, the MPC optimization problem is given as follows:
\begin{align}%{4}
 \min_{\bar{u}_0,\cdots,\bar{u}_{N-1}} && & \|\bar{x}_{N|t}\|_P + \sum_{k=0}^{N-1} \|\bar{x}_{t+k|t}\|_Q + \|\bar{u}_{t+k|t}\|_R  \nonumber \\
\text{s.t.}&& & \bar{x}_{t+k+1|t}=\tilde{A}\bar{x}_{t+k|t} + \tilde{B}\bar{u}_{t+k|t} \nonumber\\
&& & \bar{x}_{t+k|t} \in\tilde{\mathcal{X}}_{t+k|t},\forall k= 1,\dots,N 
   				   \label{eq_stochasticMpcProblemSimple}\\ 
&& & \bar{u}_{t+k|t} \in\tilde{\mathcal{U}}_{t+k|t},\forall k= 0,\dots,N-1  \nonumber
\end{align} 
%The simplifications result in the optimization problem
%\begin{align}%{4}
% \min_{u_0,\cdots,u_{N-1}} && & E\left[\|x_{N|t}\|_P + \sum_k^{N-1} \|x_{t+k|t}\|_Q + \|u_{t+k|t}\|_R\right]   \nonumber \\
%\text{s.t.}&& & x_{t+k+1|t}=\tilde{A}\tilde{x}_{t+k+1|t} + %\tilde{B}\tilde{u}_{t+k|t} \nonumber\\
%&& & x_{t+k|t} \in\tilde{\mathcal{X}}_{t+k|t},\forall k= 1,\dots,N 
%   				   \label{eq_stochasticMpcProblemSimple}\\ 
%&& & u_{t+k|t} \in\tilde{\mathcal{U}}_{t+k|t},\forall k= 1,\dots,N  \nonumber
%\end{align} 
where the predictor model with $\tilde{A}$, $\tilde{B}$ is obtained from discretizing   \eqref{eq_augmentedStateSpaceEq}.

\subsection{Algorithm}
\label{subsec_algorithm}
The proposed GP-based MPC for quadcopters is finally implemented as follows and sketched in Algorithm \ref{label_pseudocode_AlgorithmOfSMPC}.
The quadcopter first flies based on a nominal MPC while the disturbance data are collected  
%and collects disturbance data 
for the GPs as described in section \ref{subsec_disturbanceExtraction}. The MPC is updated at intervals of $T=\nobreak\SI{100}{\milli\second}$; between the MPC updates, the GP hyperparameters are trained  based on the available data as described in section \ref{section_GPsIntro}.
%The GP hyperparameters will be trained in between the MPC updates based on the available data as described in section \ref{section_GPsIntro}. 
%At the end of each time interval, it is not guaranteed that the GP hyperparameters already converged to local optima. 
At the end of each time interval, it is not guaranteed that the GP hyperparameters will converge to local optima. However, assuming that the disturbance dynamics do not change arbitrarily fast, they likely converge after certain number of iterations as was demonstrated in \cite{Solin2018}.
% A similar approach has been considered in \cite{Solin2018}. %This has similarly been done by Solin et al. \cite{Solin2018}. 

Once the GPs experienced "sufficient" training, the algorithm switches from the nominal MPC to the GP-based MPC, thus including the respective disturbance predictions. The term "sufficient" is still an open research question. In the current implementation, the switching simply happens after the GP hyperparameters have been updated at least $50$ times using \eqref{eq_hyperparametersUpdate}. In our experience, the hyperparameters typically converged in simulation flights at this point.

\begin{algorithm}
\SetKwInput{KwInput}{Input}                % Set the Input
\DontPrintSemicolon
  \textbf{Input: } $P$,$Q$,$R$,$\mathcal{X}$,$\mathcal{U}$,$p_x$,$p_u$,$N$,$\Delta T$\;
  $t=0$\;
  Perform a takeoff\;
   \While{true}
   {
   		$x_{t|t} \gets$ Telemetry\;
   		\If{\textrm{GP hyperparameters updated $50$ times}}
   		{
   		   	$\tilde{\mathcal{X}}_{t+1:N|t}, \tilde{\mathcal{U}}_{t+1:N|t} \gets $ eq. \ref{eq_tightenedConstraints}\; 
   			$u^*_{t:(t+N)|t} \gets$  eq. \ref{eq_stochasticMpcProblemSimple} \;
   		}
   		\Else
   		{
   			$u^*_{t:(t+N)|t} \gets$ eq. \ref{eq_nominalMpcProblem} \;
   		}
   		\textrm{Quadcopter} $\gets u^*_{t|t} + \begin{pmatrix}
   		u_{hover} & 0 & 0 & 0
   		\end{pmatrix}^{\top}$\;
   		\If{$t>0$}
   		{
   		$\hat{x}_{t|t-1} \gets \hat{f}(x_{t-1}, u_{t-1})$\;
   		$d_{t-1|t} \gets  \frac{x_{t} - \hat{x}_{t|t-1}}{\Delta T}$\;
		}   		
	 	\While{Within MPC update interval $\Delta T$}{
    		Update GPs with collected $d_{0:t}$\;
		}
	 	%$x_{t-1} \gets x_{t}$\;
	 	%$u_{t-1} \gets u^*_{t|t}$\;
		$t\gets t+1$\;
   }
\caption{GP-based MPC with online disturbance prediction.}
\label{label_pseudocode_AlgorithmOfSMPC}
\end{algorithm}

\section{IMPLEMENTATION AND RESULTS}
\label{sec:implementationResults}

In this section, we discuss the implementation of 
Algorithm~\ref{label_pseudocode_AlgorithmOfSMPC} 
and we demonstrate the results.
The algorithm has been implemented in C++ on a Raspberry Pi 4 B in order to evaluate the computation times. The same implementation has then been tested in simulation as an offboard controller for the PX4 autopilot using the jMAVSim simulation environment. % \cite{Meier2015}. 
The used quadcopter parameters are depicted in table \ref{label_Implementation_TableParametersOfQuadcopter}, the thrust-input necessary during hovering $u_{hover}=0.3$ and the MPC parameters are given in Table \ref{label_MPC_Params}. %\textcolor{orange}{as follows} \textcolor{red}{lets make the mpc parameters a table to improve strange spacing on left half of this page?}:
%\textcolor{orange}{
%\begin{align*}
%N&=25,\\
%Q&=\mathrm{diag}(6,6,6,6,6,6,1,1,1,1,1,1),\\
%R&=10^3 \cdot \mathrm{diag}(5,40,40,40),\\
%G^x&=\begin{pmatrix}I_{12} & -I_{12}\end{pmatrix}, \\
%h^x&= \begin{pmatrix} h_{x1} & h_{x2} & h_{x3}\end{pmatrix}^\top \\
%h_{x1} &=\begin{pmatrix}15&15&6&10&10&10&\frac{\pi}{3}&\frac{\pi}{3}\end{pmatrix},\\
%h_{x2} &=\begin{pmatrix}\frac{\pi}{5}& 2\pi&2\pi&2\pi&15&15&6&10\end{pmatrix},\\
%h_{x3} &=\begin{pmatrix}10&10& \frac{\pi}{3}&\frac{\pi}{3}&\frac{\pi}{5}&2\pi&2\pi&2\pi\end{pmatrix},\\
%G^u&=\begin{pmatrix}I_4 & -I_4\end{pmatrix}^\top, \\
%h^u&=\begin{pmatrix}1-u_{hover} & 1 & 1 & 1 & u_{hover} & 1 & 1 & 1\end{pmatrix}.
%\end{align*}}
\begin{table}[!b]
\centering
\caption{MPC Parameters}
%\textcolor{green}{
\begin{tabular}{c||c}
\hline 
N & $25$\\
\hline 
Q & $\mathrm{diag}(6,6,6,6,6,6,1,1,1,1,1,1)$\\
\hline
R & $10^3 \cdot \mathrm{diag}(5,40,40,40)$\\
\hline 
$H^x$ & $\begin{pmatrix}I_{12} & -I_{12}\end{pmatrix}$ \\
\hline 
%$h^x$ &  $\begin{pmatrix} h_{x1} & h_{x2} & h_{x3}\end{pmatrix}^\top $\\
$h^x$ &  $\begin{pmatrix} h_{x1} & h_{x2} & h_{x1} & h_{x2}\end{pmatrix}^\top $\\
\hline 
$h_{x1}$  &$\begin{pmatrix}15&15&6&10&10&10\end{pmatrix}$\\
%$h_{x1}$  &$\begin{pmatrix}15&15&6&10&10&10&\frac{\pi}{3}&\frac{\pi}{3}\end{pmatrix}$\\
\hline 
$h_{x2}$  & $\begin{pmatrix}\frac{\pi}{3}&\frac{\pi}{3}&\frac{\pi}{5}& 2\pi&2\pi&2\pi\end{pmatrix}$\\
%\hline 
%$h_{x3}$  & $\begin{pmatrix}10&10&\frac{\pi}{3}&\frac{\pi}{3}&\frac{\pi}{5}& 2\pi&2\pi&2\pi\end{pmatrix}$\\
\hline 
$H^u$ & $\begin{pmatrix}I_4 & -I_4\end{pmatrix}^\top$\\ 
\hline
$h^u$ & $\begin{pmatrix}1-u_{hover} & 1 & 1 & 1 & u_{hover} & 1 & 1 & 1\end{pmatrix}$\\
\hline 
\end{tabular} 
%}
\label{label_MPC_Params}
%\vspace{-4mm}
\end{table}%}

The LQR in \eqref{eq_dual_control_input} uses the same matrices $Q, R$ for its cost function as the MPC. The MPC's predictor model has been discretized at $\SI{10}{\hertz}$ while it has been recomputed with a rate of $\SI{30}{\hertz}$, which rendered a better flight performance during the simulation. The optimization problem has been solved using the C++-OSQP library \cite{osqp} and the osqp-eigen interface\footnote{ \url{https://github.com/robotology/osqp-eigen}, \url{http://eigen.tuxfamily.org}}. In order to connect to the PX4 Autopilot firmware and send offboard commands from the C++ script, the MAVSDK-library\footnote{\url{https://mavsdk.mavlink.io/main/en/index.html}} has been used. 

%Furthermore, a script on github from Takamasa Horibe has been used to solve discrete Algebraic Riccati equations for the LQR. <- Note necessarily used anymore.

The squared exponential kernel  has been considered for all the GPs, which are computed as LTI systems. Therefore,  Fourier-Transform of the squared exponential function has been brought into rational form via a sixth-order Taylor approximation as proposed in \cite{Hartikainen2010}. The GP hyperparameters are continuously updated as described in section \ref{subsec_algorithm} based on the last $50$ recorded data samples and a learning rate of $\eta = \begin{pmatrix} 0.03&0.01&0.005 \end{pmatrix}^\top$, see \eqref{eq_hyperparametersUpdate}. 

%The implementation has been tested in the simulation environment jMavSim (should I put in here how I edited the parameters in the simulation file??) as well as in outdoor flight tests.

\begin{figure}[htb]
\begin{tikzpicture}
    \pgfplotsset{footnotesize,samples=10}
    \begin{groupplot}[group style = {group size = 1 by 3, horizontal sep = 50pt, vertical sep = 0pt}, width = \linewidth, height = 2.7cm]
        \nextgroupplot[xticklabel={\empty},xtick={\empty},tick pos=left,ylabel={x ($m$)},
            legend style = { column sep = 10pt, legend columns = -1, legend to name = grouplegend}]
            \addplot[line width=0.5mm, color=black,densely dotted] coordinates {
						(1.0000, 0.0000)(40.0000, 0.0000)
					};\addlegendentry{Reference}%
					\addplot[line width=.5mm, color=red,dashed] coordinates {
(0.0000, -1.4281)(0.3008, -1.3277)(0.6015, -1.2566)(0.9023, -1.2597)(1.2030, -1.2982)(1.5038, -1.3320)(1.8045, -1.3731)(2.1053, -1.4316)(2.4060, -1.4882)(2.7068, -1.5128)(3.0075, -1.5638)(3.3083, -1.6888)(3.6090, -1.8281)(3.9098, -1.9459)(4.2105, -1.9995)(4.5113, -2.0255)(4.8120, -2.0144)(5.1128, -1.9811)(5.4135, -1.9339)(5.7143, -1.8854)(6.0150, -1.8402)(6.3158, -1.8108)(6.6165, -1.7865)(6.9173, -1.7630)(7.2180, -1.7289)(7.5188, -1.6919)(7.8195, -1.6423)(8.1203, -1.5957)(8.4211, -1.5448)(8.7218, -1.4983)(9.0226, -1.4542)(9.3233, -1.4322)(9.6241, -1.4312)(9.9248, -1.4457)(10.2256, -1.4754)(10.5263, -1.5064)(10.8271, -1.5252)(11.1278, -1.5394)(11.4286, -1.5494)(11.7293, -1.5534)(12.0301, -1.5536)(12.3308, -1.5555)(12.6316, -1.5532)(12.9323, -1.5544)(13.2331, -1.5614)(13.5338, -1.5731)(13.8346, -1.5831)(14.1353, -1.5919)(14.4361, -1.6064)(14.7368, -1.6282)(15.0376, -1.6513)(15.3383, -1.6703)(15.6391, -1.6865)(15.9398, -1.7018)(16.2406, -1.7159)(16.5414, -1.7257)(16.8421, -1.7259)(17.1429, -1.7162)(17.4436, -1.7130)(17.7444, -1.7109)(18.0451, -1.7103)(18.3459, -1.7125)(18.6466, -1.7129)(18.9474, -1.7088)(19.2481, -1.7076)(19.5489, -1.7117)(19.8496, -1.7142)(20.1504, -1.7182)(20.4511, -1.7228)(20.7519, -1.7206)(21.0526, -1.7166)(21.3534, -1.7052)(21.6541, -1.6832)(21.9549, -1.6505)(22.2556, -1.6149)(22.5564, -1.5760)(22.8571, -1.5383)(23.1579, -1.4908)(23.4586, -1.4487)(23.7594, -1.4130)(24.0602, -1.3821)(24.3609, -1.3601)(24.6617, -1.3459)(24.9624, -1.3357)(25.2632, -1.3305)(25.5639, -1.3343)(25.8647, -1.3510)(26.1654, -1.3801)(26.4662, -1.4073)(26.7669, -1.4279)(27.0677, -1.4449)(27.3684, -1.4588)(27.6692, -1.4716)(27.9699, -1.4897)(28.2707, -1.5063)(28.5714, -1.5203)(28.8722, -1.5371)(29.1729, -1.5572)(29.4737, -1.5712)(29.7744, -1.5738)(30.0752, -1.5758)(30.3759, -1.5760)(30.6767, -1.5749)(30.9774, -1.5833)(31.2782, -1.5991)(31.5789, -1.6179)(31.8797, -1.6368)(32.1805, -1.6612)(32.4812, -1.6938)(32.7820, -1.7340)(33.0827, -1.7686)(33.3835, -1.7955)(33.6842, -1.8107)(33.9850, -1.8137)(34.2857, -1.8186)(34.5865, -1.8223)(34.8872, -1.8212)(35.1880, -1.8085)(35.4887, -1.7854)(35.7895, -1.7585)(36.0902, -1.7397)(36.3910, -1.7308)(36.6917, -1.7313)(36.9925, -1.7351)(37.2932, -1.7355)(37.5940, -1.7235)(37.8947, -1.7049)(38.1955, -1.6880)(38.4962, -1.6793)(38.7970, -1.6875)(39.0977, -1.7200)(39.3985, -1.7558)(39.6992, -1.7953)(40.0000, -1.8277)
					}; \addlegendentry{MPC}
					\addplot[line width=.5mm, color=blue] coordinates {
(0.0000, -2.2594)(0.3008, -2.1172)(0.6015, -1.9591)(0.9023, -1.7782)(1.2030, -1.6142)(1.5038, -1.4692)(1.8045, -1.3789)(2.1053, -1.2785)(2.4060, -1.1004)(2.7068, -0.8919)(3.0075, -0.6785)(3.3083, -0.5657)(3.6090, -0.5382)(3.9098, -0.5110)(4.2105, -0.4195)(4.5113, -0.2908)(4.8120, -0.2017)(5.1128, -0.1652)(5.4135, -0.1588)(5.7143, -0.1326)(6.0150, -0.0708)(6.3158, -0.0237)(6.6165, -0.0261)(6.9173, -0.0351)(7.2180, 0.0164)(7.5188, 0.1313)(7.8195, 0.2427)(8.1203, 0.2561)(8.4211, 0.2029)(8.7218, 0.1231)(9.0226, 0.0631)(9.3233, 0.0325)(9.6241, -0.0155)(9.9248, -0.1165)(10.2256, -0.2009)(10.5263, -0.2391)(10.8271, -0.2370)(11.1278, -0.2374)(11.4286, -0.2054)(11.7293, -0.1265)(12.0301, -0.0338)(12.3308, 0.0181)(12.6316, 0.0399)(12.9323, 0.0794)(13.2331, 0.1172)(13.5338, 0.1254)(13.8346, 0.1168)(14.1353, 0.0989)(14.4361, 0.0748)(14.7368, 0.0563)(15.0376, 0.0344)(15.3383, 0.0149)(15.6391, 0.0051)(15.9398, -0.0049)(16.2406, -0.0206)(16.5414, -0.0236)(16.8421, -0.0167)(17.1429, -0.0189)(17.4436, -0.0309)(17.7444, -0.0449)(18.0451, -0.0548)(18.3459, -0.0541)(18.6466, -0.0370)(18.9474, -0.0166)(19.2481, -0.0072)(19.5489, -0.0121)(19.8496, -0.0151)(20.1504, -0.0022)(20.4511, 0.0246)(20.7519, 0.0246)(21.0526, -0.0296)(21.3534, -0.0888)(21.6541, -0.0736)(21.9549, 0.0224)(22.2556, 0.1315)(22.5564, 0.1782)(22.8571, 0.1354)(23.1579, 0.0410)(23.4586, 0.0058)(23.7594, 0.0516)(24.0602, 0.1146)(24.3609, 0.1385)(24.6617, 0.1317)(24.9624, 0.1550)(25.2632, 0.2383)(25.5639, 0.3262)(25.8647, 0.3370)(26.1654, 0.2764)(26.4662, 0.2226)(26.7669, 0.2211)(27.0677, 0.2234)(27.3684, 0.1686)(27.6692, 0.0966)(27.9699, 0.0502)(28.2707, 0.0272)(28.5714, -0.0336)(28.8722, -0.1576)(29.1729, -0.2402)(29.4737, -0.2359)(29.7744, -0.1798)(30.0752, -0.1526)(30.3759, -0.1794)(30.6767, -0.1823)(30.9774, -0.1309)(31.2782, -0.0935)(31.5789, -0.0754)(31.8797, -0.0707)(32.1805, -0.0734)(32.4812, -0.0809)(32.7820, -0.0788)(33.0827, -0.0600)(33.3835, -0.0452)(33.6842, -0.0430)(33.9850, -0.0399)(34.2857, -0.0177)(34.5865, 0.0090)(34.8872, 0.0473)(35.1880, 0.0768)(35.4887, 0.0888)(35.7895, 0.0946)(36.0902, 0.1109)(36.3910, 0.1326)(36.6917, 0.1320)(36.9925, 0.1083)(37.2932, 0.0769)(37.5940, 0.0369)(37.8947, 0.0130)(38.1955, 0.0111)(38.4962, 0.0212)(38.7970, 0.0310)(39.0977, 0.0335)(39.3985, 0.0195)(39.6992, -0.0080)(40.0000, -0.0370)
					};\addlegendentry{GP-based MPC}
        \nextgroupplot[xticklabel={\empty},xtick={\empty},tick pos=left,ylabel={y($m$)},]
           \addplot[line width=0.5mm, color=black,densely dotted] coordinates {
(0.0000, -3.0000)(0.3008, -3.0000)(0.6015, -3.0000)(0.9023, -3.0000)(1.2030, -3.0000)(1.5038, -3.0000)(1.8045, -3.0000)(2.1053, -3.0000)(2.4060, -3.0000)(2.7068, -3.0000)(3.0075, -3.0000)(3.3083, -3.0000)(3.6090, -3.0000)(3.9098, -3.0000)(4.2105, -3.0000)(4.5113, -3.0000)(4.8120, -3.0000)(5.1128, -3.0000)(5.4135, -3.0000)(5.7143, -3.0000)(6.0150, -3.0000)(6.3158, -3.0000)(6.6165, -3.0000)(6.9173, -3.0000)(7.2180, -3.0000)(7.5188, -3.0000)(7.8195, -3.0000)(8.1203, -3.0000)(8.4211, -3.0000)(8.7218, -3.0000)(9.0226, -3.0000)(9.3233, -3.0000)(9.6241, -3.0000)(9.9248, -3.0000)(10.2256, -3.0000)(10.5263, -3.0000)(10.8271, -3.0000)(11.1278, -3.0000)(11.4286, -3.0000)(11.7293, -3.0000)(12.0301, -3.0000)(12.3308, -3.0000)(12.6316, -3.0000)(12.9323, -3.0000)(13.2331, -3.0000)(13.5338, -3.0000)(13.8346, -3.0000)(14.1353, -3.0000)(14.4361, -3.0000)(14.7368, -3.0000)(15.0376, -3.0000)(15.3383, -3.0000)(15.6391, -3.0000)(15.9398, -3.0000)(16.2406, -3.0000)(16.5414, -3.0000)(16.8421, -3.0000)(17.1429, -3.0000)(17.4436, -3.0000)(17.7444, -3.0000)(18.0451, -3.0000)(18.3459, -3.0000)(18.6466, -3.0000)(18.9474, -3.0000)(19.2481, -3.0000)(19.5489, -3.0000)(19.8496, -3.0000)(20.1504, 0.0000)(20.4511, 0.0000)(20.7519, 0.0000)(21.0526, 0.0000)(21.3534, 0.0000)(21.6541, 0.0000)(21.9549, 0.0000)(22.2556, 0.0000)(22.5564, 0.0000)(22.8571, 0.0000)(23.1579, 0.0000)(23.4586, 0.0000)(23.7594, 0.0000)(24.0602, 0.0000)(24.3609, 0.0000)(24.6617, 0.0000)(24.9624, 0.0000)(25.2632, 0.0000)(25.5639, 0.0000)(25.8647, 0.0000)(26.1654, 0.0000)(26.4662, 0.0000)(26.7669, 0.0000)(27.0677, 0.0000)(27.3684, 0.0000)(27.6692, 0.0000)(27.9699, 0.0000)(28.2707, 0.0000)(28.5714, 0.0000)(28.8722, 0.0000)(29.1729, 0.0000)(29.4737, 0.0000)(29.7744, 0.0000)(30.0752, 0.0000)(30.3759, 0.0000)(30.6767, 0.0000)(30.9774, 0.0000)(31.2782, 0.0000)(31.5789, 0.0000)(31.8797, 0.0000)(32.1805, 0.0000)(32.4812, 0.0000)(32.7820, 0.0000)(33.0827, 0.0000)(33.3835, 0.0000)(33.6842, 0.0000)(33.9850, 0.0000)(34.2857, 0.0000)(34.5865, 0.0000)(34.8872, 0.0000)(35.1880, 0.0000)(35.4887, 0.0000)(35.7895, 0.0000)(36.0902, 0.0000)(36.3910, 0.0000)(36.6917, 0.0000)(36.9925, 0.0000)(37.2932, 0.0000)(37.5940, 0.0000)(37.8947, 0.0000)(38.1955, 0.0000)(38.4962, 0.0000)(38.7970, 0.0000)(39.0977, 0.0000)(39.3985, 0.0000)(39.6992, 0.0000)(40.0000, 0.0000)
					};
					\addplot[line width=.5mm, color=red,dashed] coordinates {
(0.0000, -1.3321)(0.3008, -1.2312)(0.6015, -1.1670)(0.9023, -1.2257)(1.2030, -1.4529)(1.5038, -1.8814)(1.8045, -2.5676)(2.1053, -3.2835)(2.4060, -4.1067)(2.7068, -4.7906)(3.0075, -5.3376)(3.3083, -5.7130)(3.6090, -5.8615)(3.9098, -5.8597)(4.2105, -5.7462)(4.5113, -5.5533)(4.8120, -5.3499)(5.1128, -5.1416)(5.4135, -4.9347)(5.7143, -4.7771)(6.0150, -4.6494)(6.3158, -4.5819)(6.6165, -4.5422)(6.9173, -4.5194)(7.2180, -4.4988)(7.5188, -4.4806)(7.8195, -4.4596)(8.1203, -4.4398)(8.4211, -4.4165)(8.7218, -4.3935)(9.0226, -4.3713)(9.3233, -4.3625)(9.6241, -4.3705)(9.9248, -4.3875)(10.2256, -4.4066)(10.5263, -4.4156)(10.8271, -4.4148)(11.1278, -4.4108)(11.4286, -4.4091)(11.7293, -4.4058)(12.0301, -4.3975)(12.3308, -4.3923)(12.6316, -4.3939)(12.9323, -4.4024)(13.2331, -4.4195)(13.5338, -4.4404)(13.8346, -4.4568)(14.1353, -4.4718)(14.4361, -4.4905)(14.7368, -4.5166)(15.0376, -4.5471)(15.3383, -4.5731)(15.6391, -4.5937)(15.9398, -4.6072)(16.2406, -4.6173)(16.5414, -4.6201)(16.8421, -4.6124)(17.1429, -4.5926)(17.4436, -4.5783)(17.7444, -4.5670)(18.0451, -4.5598)(18.3459, -4.5525)(18.6466, -4.5422)(18.9474, -4.5274)(19.2481, -4.5131)(19.5489, -4.5066)(19.8496, -4.4854)(20.1504, -4.3924)(20.4511, -4.1779)(20.7519, -3.8601)(21.0526, -3.4964)(21.3534, -3.1466)(21.6541, -2.7637)(21.9549, -2.3910)(22.2556, -2.0780)(22.5564, -1.8301)(22.8571, -1.6583)(23.1579, -1.5154)(23.4586, -1.4194)(23.7594, -1.3615)(24.0602, -1.3334)(24.3609, -1.3294)(24.6617, -1.3310)(24.9624, -1.3345)(25.2632, -1.3374)(25.5639, -1.3447)(25.8647, -1.3631)(26.1654, -1.3897)(26.4662, -1.4096)(26.7669, -1.4214)(27.0677, -1.4317)(27.3684, -1.4361)(27.6692, -1.4364)(27.9699, -1.4395)(28.2707, -1.4424)(28.5714, -1.4420)(28.8722, -1.4454)(29.1729, -1.4519)(29.4737, -1.4522)(29.7744, -1.4434)(30.0752, -1.4339)(30.3759, -1.4248)(30.6767, -1.4198)(30.9774, -1.4275)(31.2782, -1.4458)(31.5789, -1.4660)(31.8797, -1.4842)(32.1805, -1.5089)(32.4812, -1.5421)(32.7820, -1.5797)(33.0827, -1.6104)(33.3835, -1.6330)(33.6842, -1.6492)(33.9850, -1.6591)(34.2857, -1.6714)(34.5865, -1.6825)(34.8872, -1.6870)(35.1880, -1.6818)(35.4887, -1.6683)(35.7895, -1.6502)(36.0902, -1.6372)(36.3910, -1.6320)(36.6917, -1.6289)(36.9925, -1.6244)(37.2932, -1.6132)(37.5940, -1.5875)(37.8947, -1.5563)(38.1955, -1.5271)(38.4962, -1.5086)(38.7970, -1.5094)(39.0977, -1.5320)(39.3985, -1.5605)(39.6992, -1.5895)(40.0000, -1.6130)
					};
					\addplot[line width=.5mm, color=blue] coordinates {
(0.0000, -2.1502)(0.3008, -2.0012)(0.6015, -1.8620)(0.9023, -1.7899)(1.2030, -1.8755)(1.5038, -2.2198)(1.8045, -2.7874)(2.1053, -3.4267)(2.4060, -4.1695)(2.7068, -4.7960)(3.0075, -5.3391)(3.3083, -5.5870)(3.6090, -5.5286)(3.9098, -5.2102)(4.2105, -4.6932)(4.5113, -4.1337)(4.8120, -3.6799)(5.1128, -3.3173)(5.4135, -3.1131)(5.7143, -3.0355)(6.0150, -3.0033)(6.3158, -2.9800)(6.6165, -2.9807)(6.9173, -3.0566)(7.2180, -3.2111)(7.5188, -3.3722)(7.8195, -3.4692)(8.1203, -3.4674)(8.4211, -3.4227)(8.7218, -3.3905)(9.0226, -3.4111)(9.3233, -3.4333)(9.6241, -3.3711)(9.9248, -3.2007)(10.2256, -3.0123)(10.5263, -2.8416)(10.8271, -2.7687)(11.1278, -2.7591)(11.4286, -2.7527)(11.7293, -2.7500)(12.0301, -2.8010)(12.3308, -2.9068)(12.6316, -3.0077)(12.9323, -3.0496)(13.2331, -3.0374)(13.5338, -3.0159)(13.8346, -2.9995)(14.1353, -2.9858)(14.4361, -2.9736)(14.7368, -2.9583)(15.0376, -2.9473)(15.3383, -2.9484)(15.6391, -2.9486)(15.9398, -2.9518)(16.2406, -2.9575)(16.5414, -2.9528)(16.8421, -2.9422)(17.1429, -2.9424)(17.4436, -2.9577)(17.7444, -2.9789)(18.0451, -2.9947)(18.3459, -2.9997)(18.6466, -2.9876)(18.9474, -2.9675)(19.2481, -2.9604)(19.5489, -2.9666)(19.8496, -2.9566)(20.1504, -2.8735)(20.4511, -2.6360)(20.7519, -2.2003)(21.0526, -1.6545)(21.3534, -1.0995)(21.6541, -0.6069)(21.9549, -0.2923)(22.2556, -0.1486)(22.5564, -0.1169)(22.8571, -0.1045)(23.1579, -0.0392)(23.4586, 0.0522)(23.7594, 0.0967)(24.0602, 0.0346)(24.3609, -0.0742)(24.6617, -0.1534)(24.9624, -0.1701)(25.2632, -0.1698)(25.5639, -0.2190)(25.8647, -0.3527)(26.1654, -0.4648)(26.4662, -0.4913)(26.7669, -0.4090)(27.0677, -0.2981)(27.3684, -0.2238)(27.6692, -0.2148)(27.9699, -0.1979)(28.2707, -0.1218)(28.5714, -0.0154)(28.8722, 0.0633)(29.1729, 0.0549)(29.4737, 0.0307)(29.7744, 0.0624)(30.0752, 0.1776)(30.3759, 0.3009)(30.6767, 0.3849)(30.9774, 0.3846)(31.2782, 0.3256)(31.5789, 0.2612)(31.8797, 0.2098)(32.1805, 0.1467)(32.4812, 0.0830)(32.7820, 0.0408)(33.0827, 0.0236)(33.3835, 0.0067)(33.6842, -0.0284)(33.9850, -0.0655)(34.2857, -0.0799)(34.5865, -0.0843)(34.8872, -0.0807)(35.1880, -0.0778)(35.4887, -0.0814)(35.7895, -0.0815)(36.0902, -0.0651)(36.3910, -0.0326)(36.6917, -0.0171)(36.9925, -0.0242)(37.2932, -0.0343)(37.5940, -0.0328)(37.8947, -0.0141)(38.1955, 0.0150)(38.4962, 0.0435)(38.7970, 0.0714)(39.0977, 0.0902)(39.3985, 0.0879)(39.6992, 0.0662)(40.0000, 0.0478)
					};
        \nextgroupplot[xlabel={time ($s$)},ylabel={z ($m$)},tick pos=left] 
            \addplot[line width=0.5mm, color=black,densely dotted] coordinates {
						(1.0000, 3.0000)(40.0000, 3.0000)
					};
					\addplot[line width=.5mm, color=red,dashed] coordinates {
(0.0000, 1.9939)(0.3008, 2.1836)(0.6015, 2.3807)(0.9023, 2.5628)(1.2030, 2.6942)(1.5038, 2.7733)(1.8045, 2.8278)(2.1053, 2.8575)(2.4060, 2.8618)(2.7068, 2.8301)(3.0075, 2.7686)(3.3083, 2.6866)(3.6090, 2.6461)(3.9098, 2.6448)(4.2105, 2.6610)(4.5113, 2.6897)(4.8120, 2.7204)(5.1128, 2.7523)(5.4135, 2.7868)(5.7143, 2.8150)(6.0150, 2.8413)(6.3158, 2.8582)(6.6165, 2.8689)(6.9173, 2.8733)(7.2180, 2.8752)(7.5188, 2.8758)(7.8195, 2.8777)(8.1203, 2.8804)(8.4211, 2.8846)(8.7218, 2.8887)(9.0226, 2.8928)(9.3233, 2.8965)(9.6241, 2.8993)(9.9248, 2.8994)(10.2256, 2.8970)(10.5263, 2.8944)(10.8271, 2.8925)(11.1278, 2.8918)(11.4286, 2.8911)(11.7293, 2.8902)(12.0301, 2.8888)(12.3308, 2.8862)(12.6316, 2.8836)(12.9323, 2.8813)(13.2331, 2.8806)(13.5338, 2.8806)(13.8346, 2.8778)(14.1353, 2.8721)(14.4361, 2.8666)(14.7368, 2.8628)(15.0376, 2.8583)(15.3383, 2.8550)(15.6391, 2.8512)(15.9398, 2.8485)(16.2406, 2.8482)(16.5414, 2.8487)(16.8421, 2.8482)(17.1429, 2.8490)(17.4436, 2.8504)(17.7444, 2.8503)(18.0451, 2.8494)(18.3459, 2.8490)(18.6466, 2.8505)(18.9474, 2.8527)(19.2481, 2.8562)(19.5489, 2.8599)(19.8496, 2.8586)(20.1504, 2.8439)(20.4511, 2.8224)(20.7519, 2.8098)(21.0526, 2.8077)(21.3534, 2.8136)(21.6541, 2.8244)(21.9549, 2.8396)(22.2556, 2.8557)(22.5564, 2.8726)(22.8571, 2.8869)(23.1579, 2.9012)(23.4586, 2.9119)(23.7594, 2.9193)(24.0602, 2.9236)(24.3609, 2.9254)(24.6617, 2.9248)(24.9624, 2.9235)(25.2632, 2.9237)(25.5639, 2.9241)(25.8647, 2.9232)(26.1654, 2.9214)(26.4662, 2.9168)(26.7669, 2.9107)(27.0677, 2.9055)(27.3684, 2.9012)(27.6692, 2.8980)(27.9699, 2.8945)(28.2707, 2.8909)(28.5714, 2.8869)(28.8722, 2.8840)(29.1729, 2.8810)(29.4737, 2.8778)(29.7744, 2.8750)(30.0752, 2.8752)(30.3759, 2.8758)(30.6767, 2.8758)(30.9774, 2.8764)(31.2782, 2.8757)(31.5789, 2.8737)(31.8797, 2.8715)(32.1805, 2.8689)(32.4812, 2.8647)(32.7820, 2.8595)(33.0827, 2.8542)(33.3835, 2.8502)(33.6842, 2.8483)(33.9850, 2.8485)(34.2857, 2.8490)(34.5865, 2.8492)(34.8872, 2.8484)(35.1880, 2.8479)(35.4887, 2.8487)(35.7895, 2.8494)(36.0902, 2.8512)(36.3910, 2.8530)(36.6917, 2.8546)(36.9925, 2.8564)(37.2932, 2.8583)(37.5940, 2.8593)(37.8947, 2.8607)(38.1955, 2.8660)(38.4962, 2.8729)(38.7970, 2.8790)(39.0977, 2.8825)(39.3985, 2.8801)(39.6992, 2.8764)(40.0000, 2.8727)
					};
					\addplot[line width=.5mm, color=blue] coordinates {
(0.0000, 2.2284)(0.3008, 2.3032)(0.6015, 2.4001)(0.9023, 2.5247)(1.2030, 2.6452)(1.5038, 2.7629)(1.8045, 2.8382)(2.1053, 2.8612)(2.4060, 2.8300)(2.7068, 2.7720)(3.0075, 2.6874)(3.3083, 2.5897)(3.6090, 2.5243)(3.9098, 2.5162)(4.2105, 2.5816)(4.5113, 2.7001)(4.8120, 2.8331)(5.1128, 2.9720)(5.4135, 3.0556)(5.7143, 3.0479)(6.0150, 2.9520)(6.3158, 2.8151)(6.6165, 2.7079)(6.9173, 2.6819)(7.2180, 2.7219)(7.5188, 2.7981)(7.8195, 2.8791)(8.1203, 2.9277)(8.4211, 2.9393)(8.7218, 2.9212)(9.0226, 2.8792)(9.3233, 2.8363)(9.6241, 2.8028)(9.9248, 2.7874)(10.2256, 2.7903)(10.5263, 2.7996)(10.8271, 2.8185)(11.1278, 2.8448)(11.4286, 2.8496)(11.7293, 2.8454)(12.0301, 2.8621)(12.3308, 2.8897)(12.6316, 2.8929)(12.9323, 2.8714)(13.2331, 2.8592)(13.5338, 2.8561)(13.8346, 2.8576)(14.1353, 2.8596)(14.4361, 2.8617)(14.7368, 2.8632)(15.0376, 2.8672)(15.3383, 2.8717)(15.6391, 2.8742)(15.9398, 2.8772)(16.2406, 2.8798)(16.5414, 2.8795)(16.8421, 2.8795)(17.1429, 2.8815)(17.4436, 2.8824)(17.7444, 2.8794)(18.0451, 2.8741)(18.3459, 2.8669)(18.6466, 2.8587)(18.9474, 2.8543)(19.2481, 2.8548)(19.5489, 2.8571)(19.8496, 2.8531)(20.1504, 2.8245)(20.4511, 2.7575)(20.7519, 2.6867)(21.0526, 2.6771)(21.3534, 2.7346)(21.6541, 2.8306)(21.9549, 2.9177)(22.2556, 2.9753)(22.5564, 2.9917)(22.8571, 2.9639)(23.1579, 2.8828)(23.4586, 2.7876)(23.7594, 2.7126)(24.0602, 2.7026)(24.3609, 2.7598)(24.6617, 2.8399)(24.9624, 2.8944)(25.2632, 2.9015)(25.5639, 2.8893)(25.8647, 2.8809)(26.1654, 2.8797)(26.4662, 2.8577)(26.7669, 2.8219)(27.0677, 2.8224)(27.3684, 2.8786)(27.6692, 2.9275)(27.9699, 2.9357)(28.2707, 2.8905)(28.5714, 2.8344)(28.8722, 2.8029)(29.1729, 2.8084)(29.4737, 2.8184)(29.7744, 2.8200)(30.0752, 2.8201)(30.3759, 2.8356)(30.6767, 2.8615)(30.9774, 2.8833)(31.2782, 2.8969)(31.5789, 2.9022)(31.8797, 2.9030)(32.1805, 2.9027)(32.4812, 2.8994)(32.7820, 2.8935)(33.0827, 2.8836)(33.3835, 2.8787)(33.6842, 2.8744)(33.9850, 2.8649)(34.2857, 2.8526)(34.5865, 2.8439)(34.8872, 2.8407)(35.1880, 2.8412)(35.4887, 2.8437)(35.7895, 2.8448)(36.0902, 2.8453)(36.3910, 2.8505)(36.6917, 2.8609)(36.9925, 2.8737)(37.2932, 2.8772)(37.5940, 2.8734)(37.8947, 2.8658)(38.1955, 2.8590)(38.4962, 2.8560)(38.7970, 2.8579)(39.0977, 2.8614)(39.3985, 2.8662)(39.6992, 2.8695)(40.0000, 2.8677)
					};     
    \end{groupplot}
    %\node at ($(3.2,-3.6cm)$) {\ref{grouplegend}}; 
\end{tikzpicture}
\caption{Trajectory of a quadcopter using a \textcolor{red}{conventional MPC} (red, dashed) and \textcolor{blue}{GP-based MPC} (blue, solid)  under strong wind disturbances in jMAVSim. The black, dotted line represents the reference trajectory. }
\label{img_resultPlots}
%\vspace{-4mm}
\end{figure}
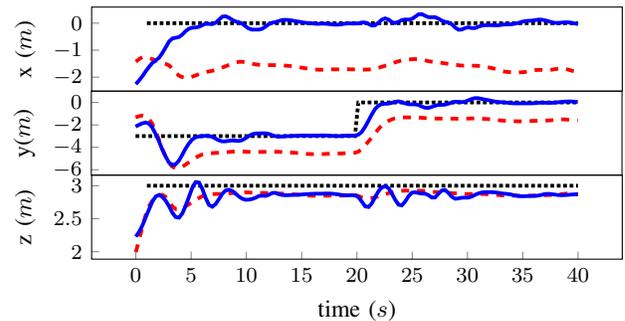

Next, the implementation has  been tested on a potent host machine in order to control a quadcopter within the jMAVSim simulation environment \cite{Meier2015} while the respective computation times that a Raspberry Pi 4 B would have required to compute the algorithm have been imitated via sleep commands. The goal was to track the reference trajectory depicted in 
Fig.~\ref{img_resultPlots}. The resulting trajectories of the simulated flights under heavy wind conditions using a nominal and the GP-based MPC are depicted in Fig.~\ref{img_resultPlots}. The jMavSim environment models wind by filtering white gaussian noise (variance set to set to $\SI{24}{\meter\per\second}$) and adding a mean wind speed vector (length set to to $\SI{22}{\meter\per\second}$) on top. However, various wind speed settings have been tested with similar results.

%{\color{red}However, various wind speeds have been tested with similar results.}
%H

\begin{table}[!h]
\centering
\caption{Parameters of the quadcopter.}
\begin{tabular}{c||c||c||c}
\hline 
$m [\SI{}{\kilo \gram}]$ & $I_{xx}[\SI{}{\kilo \gram \meter^2}]$ & $I_{yy}[\SI{}{\kilo \gram \meter^2}]$ & $I_{zz}[\SI{}{\kilo \gram \meter^2}]$   \\ 
\hline 
\hline 
$1.862$ & $0.0429$ & $0.0437$ & $0.0753$  \\
\hline 
\multicolumn{4}{c}{ }\\
%\hline \vspace{1cm}
%\hline \\ 
\hline
%\thickhline 
$k_D[\SI{}{\frac{\newton \second^2}{\meter}}]$ & $T_{max} [\SI{}{\newton}]$ & $\tau_{x,max} = \tau_{y,max}[\SI{}{\newton \meter}]$ & $\tau_{z,max}[\SI{}{\newton \meter}]$ \\ 
\hline 
\hline 
$0.1735$ & $62.06$ & $4.6548$ & $1.7$  \\
\hline 
\end{tabular} 
\label{label_Implementation_TableParametersOfQuadcopter}
%\vspace{-4mm}
\end{table}

The takeoff is performed by an automated sequence of the PX4 autopilot. The nominal and GP-based MPC take over midflight and try to track the depicted trajectory. Finally, the quadcopter lands, again using an automated sequence of the PX4 autopilot. Notably, the GP-based MPC performs significantly better in this task with a root-mean-square (RMS) euclidean distance of $\SI{0.82}{\meter}$ to the reference trajectory compared to the nominal MPC with an RMS euclidean distance of $\SI{2.83}{\meter}$ to the reference trajectory.

Furthermore, it can be seen that the GPs tend to introduce oscillations, especially in the $z$-coordinate. Interestingly, oscillations occur in the z-coordinate when the quadcopter changes its y-position. This is likely the cause of a mismatch between the nonlinear model and the true system dynamics, which overlies the disturbance estimates and may lead to undesired feedback dynamics between the GPs and the MPC. This indicates, that the proposed method relies on precise model predictions.

\input{inputSignalPlots}

These oscillations, as well as a more aggressive behaviour of the GP-based MPC compared to a conventional MPC can also be seen when analyzing the computed input signals. During the flight, the GP-based MPC computed oscillating actuation signals with a  higher amplitude (see Fig. \ref{img_resultPlotsjMAVSimInputs}).

The C++ implementation in this work proved the feasibility of the proposed control approach for real-time applications: On a Raspberry Pi 4 Model B, computing the nominal MPC took about $\SI{10}{\milli \second}$ while the computation of the MPC with GPs took about $\SI{20}{\milli \second}$. Updating the hyperparameters of a single GP with a batch size of $50$ input-output data pairs required up to $\SI{2}{\milli \second}$.


\section{CONCLUSIONS}
\label{sec:conclusions}
In this paper, the applicability of a GP-based MPC for predicting disturbances acting on a quadcopter 
and reducing their effects has been demonstrated in simulation and a practical implementation framework has been presented. It has been shown that this approach yields superior performance over a conventional MPC when disturbances affect the system and that the algorithm scheme is computable in real-time.

In order to verify the potentials that this approach shows in simulation and to prove its robustness, its algorithm will be evaluated on indoor flight tests with a focus on reproducibility. As future research directions: the computational complexity of the GP-based MPC could be further reduced by applying Infinite Horizon Gaussian Processes \cite{Solin2018}, which may also remove artefacts introduced by the truncation of the batch data.  %An interesting and very important theoretical research work may be based on finding a 
An interesting research issue is to define an
appropriate classifier that allows to determine when one should switch from the nominal to the GP-based MPC. Moreover, to deal with  model mismatch effects, the proposed GP-based MPC could be combined with a preceding learning of such mismatches as shown in \cite{Torrente2021}. A more complex quadcopter model for the MPC and disturbance estimation will be used in future investigations. 

\section*{ACKNOWLEDGMENT}

Finally, we would like to thank David Pysik, Tavia Plattenteich, and Ievgen Zhavzharov for their valuable input and support with the PX4 autopilot.
%The preferred spelling of the word ÒacknowledgmentÓ in America is without an ÒeÓ after the ÒgÓ. Avoid the stilted expression, ÒOne of us (R. B. G.) thanks . . .Ó  Instead, try ÒR. B. G. thanksÓ. Put sponsor acknowledgments in the unnumbered footnote on the first page.

%\addtolength{\textheight}{-10cm}   % This command serves to balance the column lengths

%%%%%%%%%%%%%%%%%%%%%%%%%%%%%%%%%%%%%%%%%%%%%%%%%%%%%%%%%%%%%%%%%%%%%%%%%%%%%%%%

%References are important to the reader; therefore, each citation must be complete and correct. If at all possible, references should be commonly available publications.
\addtolength{\textheight}{-1.7cm}  

\bibliographystyle{IEEEtran}
\bibliography{bib}%IEEEabrv 

\end{document}